\documentclass[aps,prd,twocolumn,groupedaddress,nofootinbib,showpacs,eqsecnum]{revtex4}

\usepackage{euscript,graphicx,amssymb,subfigure}

\usepackage{amsfonts}
\usepackage{amsmath}
\usepackage{amssymb}
\usepackage{graphicx}
\usepackage{euscript}
\usepackage{color}
\usepackage{subfigure}

\begin{document}

\title{Is the Universe Transparent?}

\author{Kai Liao\footnote{{\tt liaokai@mail.bnu.edu.cn}}}
\affiliation{School of Science, Wuhan University of Technology, Wuhan 430070, China}
\author{A. Avgoustidis\footnote{{\tt anastasios.avgoustidis@nottingham.ac.uk}}}
\affiliation{School of Physics and Astronomy, University of Nottingham, University Park, Nottingham NG7 2RD, England}
\author{Zhengxiang Li\footnote{{\tt zxli918@bnu.edu.cn}}}
\affiliation{Department of Astronomy, Beijing Normal University, Beijing 100875, China}

\begin{abstract}
We present our study on cosmic opacity, which relates to
changes in photon number as photons travel from the source to the
observer. Cosmic opacity may be caused by absorption/scattering
due to matter in the universe, or by extragalactic magnetic fields that
can turn photons into unobserved particles (e.g. light axions, chameleons,
gravitons, Kaluza-Klein modes), and it is crucial to correctly interpret
astronomical photometric measurements like type Ia supernovae
observations. On the other hand, the expansion rate at
different epochs, i.e. the observational Hubble parameter data
$H(z)$, are obtained from differential ageing of passively evolving
galaxies or from baryon acoustic oscillations and thus are not affected
by cosmic opacity. In this work, we first construct opacity-free
luminosity distances from $H(z)$ determinations, taking correlations
between different redshifts into consideration for our error analysis. Moreover,
we let the light-curve fitting parameters, accounting for distance
estimation in type Ia supernovae observations, free to ensure that our analysis
is authentically cosmological-model-independent and gives a robust result.
Any non-zero residuals between these two kinds of luminosity
distances can be deemed as an indication of the existence of cosmic
opacity. While a transparent universe is currently consistent with the data,
our results show that strong constraints on opacity (and consequently on physical
mechanisms that could cause it) can be obtained in a cosmological-model-independent fashion.
\end{abstract}


\date{\today}
\maketitle

\section{Introduction}
It was the unexpected dimming of type Ia supernovae (SNe Ia) that
revealed the accelerated expansion of the
universe\cite{Riess,Perlmutter}. Although the existence of cosmic
acceleration has been confirmed by several other independent probes,
initially there was some debate on the interpretation in terms of an
underlying physical mechanism for the observed SNe Ia dimming. For
example, soon after\cite{Riess,Perlmutter}, a cosmological distribution
of dust was proposed as an alternative explanation for the obscuration of
distant SNe Ia\cite{Agurrie1999a,Agurrie1999b}. Furthermore, cosmic
opacity may be caused by other exotic mechanisms, where
extragalactic magnetic fields turn photons into light
axions\cite{axion1,axion2,axion3}, gravitons\cite{Chen1995},
Kaluza-Klein modes associated with
extra-dimensions\cite{Deffayet2000}, or a chameleon
field\cite{Khoury2004,Burrage2008}, thus violating photon
number conservation. Indeed, the extinction effects of SNe Ia due to dust in
their host galaxies and the Milky Way have been well-modeled and they
pose no threat to the conclusion of cosmic acceleration. However,
exotic mechanisms for general cosmic opacity and the
influence on astronomical photometric measurements are still not fully
understood. Therefore, the question of whether cosmic opacity contributes
to the dimming of distant SNe Ia remains open. In other words, can cosmic
opacity mimic the behavior of dark energy to make the universe look like
accelerating at a different rate than it actually is? At the very least, this
issue is important for reliable cosmological parameter determination in the era
of precision cosmology, as opacity could be responsible for part of the observed
dimming of standard candles. Therefore, as cosmological
data precision improves, it is necessary to better quantify the transparency
of the universe and try to distinguish any relevant dimming effects.
\par
Any kind of photon number non-conservation can result in
deviations from the distance-duality relation (DDR)\cite{Etherington}:
$D_L=D_A(1+z)^2$, where $z$ is the redshift, and $D_L$, $D_A$ are
the luminosity distance and angular diameter distance, respectively.
This relation holds on three conditions\cite{Bassett}:
(1)space-time in our universe is described by Riemannian geometry;
(2)photons travel along null geodesics; (3)the number of
photons between the source and the observer is conserved. The first
two requirements are fundamental and have strong physical bases.
In contrast, the violation of the last one is certainly possible in a wide
range of well-motivated models. Recently,
a great deal of effort has been devoted in checking the validity of the
DDR with astronomical observations\cite{Uzan2004,Bernardis2006,Holanda2010,Li2011,Nair2011,Holanda2012}.
Meanwhile, there were also many studies that focused on testing
cosmic opacity under the assumption that any possible deviations from
the DDR originate from non-conservation of the number of photons between
emission at the source and detection\cite{axion2,Chen2012,Li2013,Holanda2013,Liao,Holanda2014}.
There are two general ways to carry out these studies. The first
is to confront the luminosity distances derived from SN Ia
observations with the directly measured angular diameter distances
of galaxy clusters or those inferred from baryon acoustic
oscillation observations\cite{Chen2012,Bassett2004a,Bassett2004b,Song2006,Avgoustidis2009,Li2013,Nair2012,More2009}.
On the other hand, in Refs.~\cite{Avgoustidis2009,Holanda2013,Liao,Holanda2014}, distances
derived from other opacity-independent probes, e.g. observational determinations
of the Hubble parameter $H(z)$ based on differential ageing of
passively evolving galaxies (also dubbed ``cosmic chronometers")
\cite{Jimenez2002}, were proposed to test or even quantify cosmic opacity
by comparing these distances with those from SN Ia observations.
\par
However, it is noted that the luminosity distances of SNe Ia used in
previous analyses were derived from Hubble diagrams where the
light-curve fitting parameters, accounting for distance estimation of SNe Ia,
were determined from global fitting in the context of concordance
cosmology and in this sense were
cosmological-model-dependent~\cite{Yang,Li2014}. What is more, in
Refs.~\cite{Holanda2013,Liao}, the authors constructed luminosity
distances from $H(z)$ data but did not take the correlations between
different redshifts into account. This treatment would lead to inaccurate
estimations of the errors. The two defects discussed above may give
results that are both biased by the assumed model of cosmology and
statistically incorrect. Here, in order to achieve a
reasonable and compelling test for cosmic opacity, we pay
attention to these issues by using the latest JLA SNe Ia (joint
light-curve analysis of the SDSS-II and
SNLS)~\cite{Kessler2009,Conley2011,Betoule2014} and taking
the correlations between different redshifts into
consideration in our error analysis.

This paper is organized as follows: In Section II, we introduce the
data used in our work, including SN Ia samples and $H(z)$ data. In
Section III, we discuss our improved method to constrain cosmic
opacity. In Section IV, we present our results and analysis. Finally,
we summarize our work in Section V.

\section{Data}
To achieve a cosmological-model-independent analysis, we need to select the observational
data very carefully to avoid cases where data have been obtained under a specific cosmological
model. Cosmological studies can suffer from the so-called ``circularity problem", i.e., the use of
data from a certain cosmological model to constrain another one, which can often lead to biased
or incorrect conclusions. We now introduce the SN Ia observations and Hubble parameter data
that are independent of cosmological model.

\subsection{SN Ia observations}
We adopt a joint light-curve analysis sample of SN Ia observations (JLA) obtained by the SDSS-II and
SNLS collaborations\cite{Betoule2014}.
It contains several low-redshift samples (z$<$0.1), all three seasons from the SDSS-II (0.05$<$z$<$0.4),
and three years from SNLS (0.2$<$z$<$1), totally 740 well-measured events. \par
In theory, the explosion of SN Ia has a universal physical basis, as the collapse is triggered when the
white dwarf achieves the Chandrasekhar limit. Therefore, the peak absolute magnitude $M_{max}$ is constant.
Using a Cepheid variable at the same redshift, one can know its value and the modulus or the luminosity distance
is the difference between the absolute and the observed magnitude $m_{max}$:
\begin{equation}
\mu=5\,\log D_L(Mpc)+25=m_{max}-M_{max}. \label{modulus}
\end{equation} \par
In reality, there exits a variation of $M_{max}$ related to the shape and color of
the light curve, and the $m_{max}$ is affected by extinction. Therefore, a modified version of
Eq. (\ref{modulus}) was proposed in \cite{Guy2007} known as the SALT method:
\begin{equation}
\mu_B(\alpha,\beta,M_B)=m_B-M_B+\alpha x-\beta c, \label{mub}
\end{equation}
where $m_B$ is the rest-frame peak magnitude in the B band, $x$ is the stretch determined by
the shape of the SN Ia light curve and $c$ is the color measurement\cite{Tripp1998}. Note that
 $m_B, x, c$ are all derived from the observed light curve and are thus independent of cosmological model.
 $\alpha$ and $\beta$ are nuisance parameters that characterize the stretch-luminosity
and color-luminosity relationships,
and are related to the well-known broader-brighter and bluer-brighter relationships. $M_B$
is also a nuisance parameter standing for the B band absolute magnitude.
\par
Releases of SN Ia observations are usually presented as distance modulus $\mu$ used for
cosmological study.
However, this depends on the cosmological model and the value of $H_0$. In JLA
samples\cite{Betoule2014}, the authors used flat $\Lambda CDM$ model as the standard
to minimize the $\chi^2$:
\begin{equation}
\begin{split}
&\chi^2(\alpha,\beta,M_B,\Omega_M,H_0;z)=\\
&\sum\left[\frac{\mu_B(\alpha,\beta,M_B;z)-\mu^{\Lambda CDM}(\Omega_M,H_0;z)}{\sigma_{total}}\right]^2
\end{split}
\end{equation}
where $H_0=70\, {\rm km}\, {\rm s}^{-1} {\rm Mpc}^{-1}$ was fixed. They
obtained ($\alpha, \beta, M_B$)=($0.141\pm0.006,3.101\pm0.075,-19.05\pm0.02$) including systematic errors
and ($0.140\pm0.006,3.139\pm0.072,-19.04\pm0.01$) for statistical errors only.
\par
Since it is obvious that the distance modulus depends on the cosmological model, we find
all previous studies on cosmic opacity are not cosmological-model-independent. However,
a quantification of these effects on biasing and affecting the uncertainties of opacity constraints
is currently lacking. In this work, we directly take the observational quantities ($m_B, x, c$)
and their errors ($\sigma_{m_B},\sigma_x,\sigma_c$) as our supernova data in our analysis.
We shall treat the nuisance parameters ($\alpha,\beta,M_B$) as additional parameters, uniformly distributed
over appropriate prior ranges, which can be marginalized over in a Bayesian fashion, thus resulting in a
cosmological-model-independent constraint on cosmic opacity.

\subsection{Hubble parameter data}
\begin{figure}
\includegraphics[width=9cm]{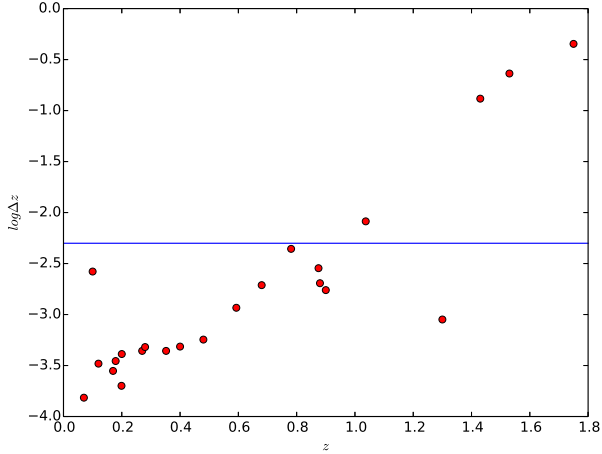}
\caption{$\Delta z$ distribution with a cutoff $\Delta z<0.005$.
        $\Delta z$ increases with $z$ since there are less observed SNe at larger redshift. }
\label{zdis}
\end{figure}

The use of observational $H(z)$ data
has successfully been an independent and powerful tool for exploring the evolution of the universe and the
role of dark energy in driving cosmic acceleration. The main advantage is that $H(z)$ data contain direct information about
the expansion of the universe at different redshifts, whereas other methods can only get cosmic
distances in the form of an integral of $H(z)$ over redshift, losing the fine structure. There are mainly
two ways to obtain $H(z)$ data. The first
one is known as ``differential ageing method" (DA), based on differential ages of red-envelope
galaxies consisting of uniform stellar populations. Subtracting the spectra between galaxies at nearby redshifts
and fitting stellar population models returns a relative age, which, given that the stellar populations in those galaxies
evolve passively, corresponds to a relative cosmic ageing. $H(z)$ is given by the following relationship:
\begin{equation}
H(z)=-\frac{1}{1+z}\frac{dz}{dt}.
\end{equation}
In this method, we assume cosmic opacity is not strongly
wavelength-dependent in the (relatively narrow) optical band and thus $H(z)$ data are opacity-free.
A discussion on the relation between
wavelength and cosmic opacity can be found in Li et. al.\cite{Li2013}.
In this work, we adopt 19 $H(z)$ data points obtained from the DA
method in \cite{Simon2005,Stern2010,Moresco2012,Zhang2012} (see also \cite{Farooq}),
which we show in Table \ref{hz}. We have excluded 4
data points that have large differences in redshift ($\Delta z>0.005$) from the
nearest observed SNe Ia. This cutoff is chosen for two reasons: firstly,
it is small enough compared to the observational errors and can be ignored,
therefore it is widely used in the literature; secondly, it allows us to include most
of the available $H(z)$ data points, see Fig. \ref{zdis}.
\par
The second way to get $H(z)$ is from from baryonic acoustic oscillations (BAO)
as a standard ruler in the radial direction, known as the ``Peak Method".
This is completely free of cosmic opacity since it is independent
of the measured flux. However, we emphasize that this method is obviously
based on the assumed cosmological model, and should therefore be abandoned
in our work where we endeavour to conduct a cosmological-model-independent analysis.

\section{Methodology}
Based on previous works\cite{Holanda2013,Liao}, we introduce an improved method to get
luminosity distances that are not affected by cosmic opacity and are also independent of any
specific cosmological model. We consider constructing 19 luminosity distances $D^H_L(z_n), n=1,2...19$,
from the $H(z)$ data at the corresponding redshifts by:
\begin{equation}
D^H_L(z_n)=c(1+z_n) \int_0^{z_n}{dz^\prime \over H(z^\prime)}
\approx {c(1+z_n)\over 2} \sum_{i=1}^{n} B_i,
\end{equation}
where
\begin{equation}
B_i=(z_i-z_{i-1})\left[ {1\over H(z_i)}+{1\over H(z_{i-1})} \right]
\end{equation}
is the $i^{th}$ bin that contributes to the integration, $H_0=H(z_0=0)=73.8 \pm 2.4\, {\rm km}/{\rm s}/{\rm Mpc}$\cite{Riess2011}
is the Hubble constant, and $c$ is the speed of light. The systematic error related to
the integral approximation has been shown to be much smaller than the observational
one\cite{Liao}.

\par

It is very important to note that these constructed luminosity distances are correlated with
each other since they have been obtained by an accumulating process over the bins $B_i$.
Therefore, we
have to calculate the $19\times19$ covariance matrix for $D_L^H(z_n)$ rather than
19 independent errors\cite{Holanda2013,Liao}. Also, even for a specific luminosity distance,
previous investigations have used an inappropriate error estimate not considering the correlations
between adjacent bins. In our analysis, we strictly follow the definition of covariance, i.e.,
$Cov(X,Y)=E(XY)-E(X)E(Y)$, where $X, Y$ are arbitrary variables, and $E$ stands for
the mathematical expectation. We only take the original $H(z)$ data points as independent measurements.
The covariance matrix $\textbf{C}^H$ for the constructed luminosity distances can be expressed as:
\begin{equation}
\begin{split}
&C_{mn}^H:= Cov\left[ D_L^H(z_m),D_L^H(z_n)\right]=\\
&\frac{c^2(1+z_m)(1+z_n)}{4} \left[ E(\sum_{i=1}^{m}\sum_{j=1}^{n}B_i B_j )-E(\sum_{i=1}^{m} B_i)E(\sum_{j=1}^{n} B_j)\right],
\end{split}
\end{equation}
where the key is to calculate $E(B_i B_j)$. Since $i, j$ are symmetric, we give its expression
for $i\leq j$:
\begin{widetext}
\begin{equation}
E(B_i B_j)=
\left\{
\begin{array}{ll}
(z_i-z_{i-1})^2\left[\frac{1}{H(z_i)^2}+\frac{1}{H(z_{i-1})^2}+
\frac{\sigma_H(z_i)^2}{H(z_i)^4}+\frac{\sigma_H(z_{i-1})^2}
{H(z_{i-1})^4}+\frac{2}{H(z_i)H(z_{i-1})}\right], & \hbox{i=j;} \\
(z_i-z_{i-1})(z_{i+1}-z_i)\left[\frac{1}{H(z_i)H(z_{i+1})}+
\frac{1}{H(z_i)^2}+\frac{\sigma_
H(z_i)^2}{H(z_i)^4}+\frac{1}{H(z_{i-1})H(z_{i+1})}+
\frac{1}{H(z_{i-1})H(z_{i})}\right], & \hbox{i+1=j;} \\
(z_i-z_{i-1})(z_j-z_{j-1})\left[\frac{1}{H(z_i)H(z_j)}+
\frac{1}{H(z_i)H(z_{j-1})}+\frac{1}{H(z_{i-1})H(z_j)}+
\frac{1}{H(z_{i-1})H(z_{j-1})}\right], & \hbox{i+1$<$j.}
  \end{array}
\right.
\end{equation}
\end{widetext}

The supernova observations are affected by cosmic opacity through an optical depth:
\begin{equation}
D^S_{L,B}(z)=10^{\mu_B(\alpha,\beta,M_B;z)/5-5}=D^S_L(z)e^{\tau(z)/2}, \label{dl}
\end{equation}
where $D^S_{L,B}$ is the observed luminosity distance from the B band and
$D^S_L$ stands for the true luminosity corresponding to SN Ia. We parameterize the optical depth
$\tau(z)$ as:
\begin{equation}
\tau(z)=2\epsilon z,
\end{equation}
since it must return to 0 for $z=0$ and, when $z$ is small, the Taylor Expansion should work.
Other parameterizations should give similar results.
\par
In order to compare the constructed luminosity distance from $H(z)$ data with that from
SN Ia observations at the same redshift, we follow Holanda et. al. 2010\cite{Holanda2010}
and Li et. al. 2011\cite{Li2011}, i.e., we find the nearest redshift to $H(z)$ data from SNe Ia. We
summarize the differences between nearest redshifts in Table~\ref{hz}, excluding $H(z)$ at
$z=1.037, 1.43, 1.53, 1.75$, which have redshift differences $\Delta z=0.0082, 0.131, 0.231, 0.451$ that
are deemed too large according to the criterion described above.

\begin{table*}
\begin{center}
\begin{tabular}{lcccccccccc}
\hline\hline
$z$ & $H(km \cdot s^{-1} \cdot Mpc^{-1})$ & $\sigma_{H}$ & $D^H_L(Mpc)$ &  $\Delta z$ & $m_B$&
$\sigma_{m_B}$ &x &$\sigma_x$&c&$\sigma_c$ \\
\tableline
0.07 &69 &19.6 &315 &0.00015&18.44505 &0.14638 &-0.43858& 0.29398 &0.01613 &0.05654\\
0.1 &69 & 12 &467   &0.00264&19.20614 &0.11332& 1.41904 &0.14596& -0.02329& 0.02300\\
0.12 &68.6 &26.2 &574 &0.00033&19.57897 &0.11217& -0.01434 &0.12814& -0.06787 &0.02216\\
0.17& 83 & 8 &833  &0.00028&20.13472&0.11204&0.86017&0.20045&-0.08973&0.02323\\
0.179& 75 &4 &880  &0.00035&20.35933&0.11332&-0.16205&0.20785&-0.10021&0.02610\\
0.199& 75 &5 &990  &0.00020&20.51988&0.12426&-0.44361&0.71455&-0.07573&0.04594\\
0.2& 72.9& 29.6 &996 &0.00041&20.84984&0.11521&1.11501&0.30976&0.01769&0.03302\\
0.27& 77& 14 &1410  &0.00044&21.67852&0.11868&-1.94823&0.77070&-0.06174&0.04803\\
0.28 &88.8 &36.6 &1468 &0.00048&21.34961&0.12190&0.69425&0.46152&-0.03017&0.04276\\
0.352 &83 &14 &1891  &0.00044&22.68491&0.08635&-0.67440&0.16430&0.03745&0.02466\\
 0.4& 95& 17 &2186  &0.00049&22.58735&0.08755&-0.72529&0.12221&-0.00466&0.02591\\
0.48&97 &62 &2681  &0.00057&22.94876&0.08805&-0.20702&0.12623&-0.02917&0.02490\\
0.593& 104 &13 &3423  &0.00117&23.38618&0.09044&1.58102&0.17420&0.02252&0.03380\\
0.68 &92& 8 &4059 &0.00194&23.49942&0.09416&0.54088&0.25811&-0.15070&0.04979\\
0.781 &105 &12 &4854 &0.00441&24.39777&0.09837&0.13522&0.28653&0.09926&0.07032\\
0.875 &125 &17 &5573 &0.00285&24.42659&0.11917&-0.74262&0.58221&-0.19630&0.05829\\
 0.88& 90& 40 &5615  &0.00203&24.32923&0.10310&1.74240&0.43591&-0.11251&0.06032\\
 0.9& 117 & 23 &5787  &0.00174&24.40844&0.11134&0.90217&0.38881&-0.15435&0.05935\\
1.3& 168& 17 &8845 &0.00089&25.69123&0.12806&0.66432&0.35756&0.00990&0.03686\\
\hline\hline
\end{tabular}
\end{center}
\caption{19 observational Hubble parameter data from ``differential
age" method. Since JLA samples consist of much more data within $z\sim1$,
the redshift differences of the nearest SNe Ia to $H(z)$ data
 are so small $\sim10^{-4}$ that they can be neglected. We also show the corresponding constructed luminosity
distances from $H(z)$ and SN Ia data at the corresponding redshifts from observed light curves:
$m_B, x, c$ and their errors.
}\label{hz}
\end{table*}

We now give the statistics for constraining cosmic opacity parametrized by $\epsilon$. We use $\chi^2$:
\begin{equation}
\chi^2=\Delta \textbf{P}^T\textbf{C}^{-1}\Delta \textbf{P},
\end{equation}
where $\Delta\textbf{P}(\alpha,\beta,M_B,\epsilon)$ is the difference between the constructed luminosity distances $D^H_L$
from the $H(z)$ data and the true luminosity distances $D^S_L$ derived from Eq. (\ref{dl}) from the SN Ia data:

\begin{equation}
\Delta\textbf{P}=\left(
                   \begin{array}{c}
                     D^S_L(z_1)-D^H_L(z_1) \\
                     D^S_L(z_2)-D^H_L(z_2) \\
                     ... \\
                     D^S_L(z_{19})-D^H_L(z_{19}) \\
                   \end{array}
                 \right).
\end{equation}
The covariance matrix \textbf{C} consists of $\textbf{C}^H$ from $H(z)$ and the errors related to
SN Ia observations $D^S_L$. The former considers the errors of the constructed luminosity distances
and their correlations, while the latter comes from the observational errors of SNe Ia only:
\begin{equation}
\textbf{C}=\textbf{C}^H+\textbf{C}^S,
\end{equation}
$\textbf{C}^S(M_B,\alpha,\beta,\epsilon)$ is the covariance matrix for SNe Ia and only the diagonal elements
for statistical uncertainties are considered since only 19 SNe Ia of 740 events are selected to match the H(z) data,
\begin{equation}
C^S_{ii}=\sigma_{D^S_L}^2(z_i)=\left[\frac{\ln10}{5}D^S_L(z_i)\right]^2\sigma^2_{\mu^S_L}(z_i),
\end{equation}
with
\begin{equation}
\sigma^2_{\mu^S_L}(z_i)=\sigma^2_{\mu_B}(z_i)=\sigma^2_{m_B}(z_i)+\alpha^2\sigma_x^2(z_i)+\beta^2\sigma_c^2(z_i).
\end{equation}
The likelihood distribution $L(\alpha,\beta,M_B,\epsilon)\propto exp(-\chi^2/2)$ and we consider uniform
distributions ($\alpha=[-1.0,1.0],\beta=[-1.0,10.0],M_B=[-21.0,-17.0],\epsilon=[-1.5,1.5]$) as our parameter priors
since we can not explore infinite ranges. We use
PyMC\footnote{{\tt http://github.com/pymc-devs/pymc}}, a python module
that implements Bayesian statistical models and fitting algorithms, including Markov chain Monte Carlo (MCMC),
to generate sample points of the probability distribution. Then, we apply a public package ``triangle.py" in
GITHUB\footnote{{\tt http://github.com/dfm/triangle.py}} to plot our constraint contours.

\section{Results and Analysis}
\begin{figure}
\includegraphics[width=9cm]{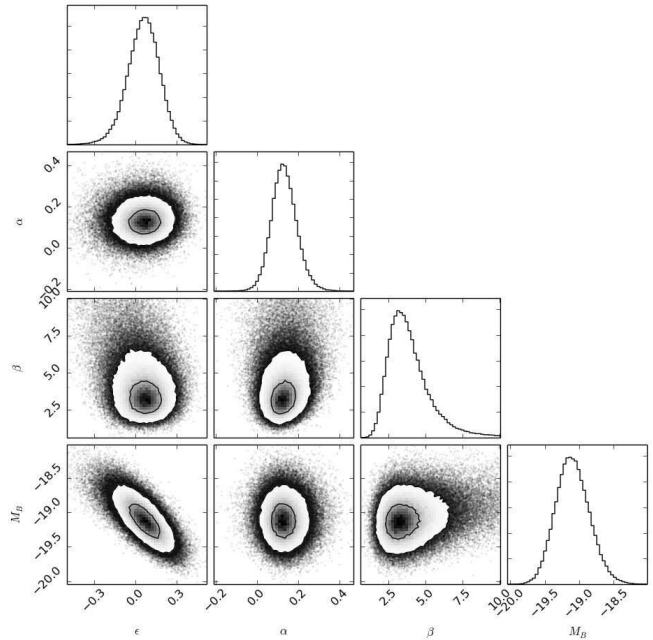}
\caption{The 1-D and 2-D marginalized distributions and $1\sigma$
and $2\sigma$ constraint contours for cosmic opacity $\epsilon$ and
SN Ia nuisance parameters ($\alpha, \beta, M_B$), respectively.
  The results are from MCMC sampling.}
\label{result}
\end{figure}

Our main results are presented in Fig. \ref{result}. We give the corner plots for $(\epsilon,\alpha,\beta,M_B)$ with
1-D distributions for each parameter and 2-D constraints for combinations of any two parameters. The inner and
outer contours stand for $1\sigma$ and $2\sigma$ ranges, respectively. We also summarize individual results
numerically in Table. \ref{stats}.
\par

Since the output of MCMC is a series of points for each parameter and the probability distribution function (PDF)
is non-Gaussian, we adopt two kinds of statistics to calculate our results and corresponding errors. The first one
is as follows: we take the value corresponding to the densest position of the distribution as the best-fit value, then
find two values on both sides of it that have the same density and contain $68.3\%$ of the distribution as
$1\sigma$ lower and upper limits, respectively. $2\sigma$ corresponds to $95.4\%$. We refer to this as ``BEST"
statistics. Moreover, we consider the mean value with errors calculated as follows: find the $68.3\%/2$ area for
both sides of the mean value, corresponding to the lower and upper limits. We call this ``MEAN" statistics.
\par
From the figure, one can easily see that $\epsilon=0$ is located near the center of our contours implying a transparent
universe is favoured by the observational data.
For the SN Ia nuisance parameters $(\alpha,\beta,M_B)$,
we compare our results with those of Betoule et. al. 2014\cite{Betoule2014}, where
($\alpha, \beta, M_B$)=($0.141\pm0.006,3.101\pm0.075,-19.05\pm0.02$) including systematic errors
and ($0.140\pm0.006,3.139\pm0.072,-19.04\pm0.01$) for statistical errors only.
We find they are sightly different but consistent with each other within $1\sigma$.
There are three distinct aspects contributing to this small difference: firstly, we only take 19 data points while Betoule et. al. 2014 used the whole JLA sample;
secondly, we have considered the impact of cosmic opacity which is degenerate with $M_B$, as can be seen from the ($\epsilon, M_B$) plane constraint in Fig. \ref{result}.
Therefore, a positive cosmic opacity will increase the intrinsic brightness such that $M_B$ is slightly smaller than that in Betoule et. al. 2014;
lastly, we use a different standard to calculate the luminosity distances of supernovae.
\par
\begin{figure}
\includegraphics[width=9cm]{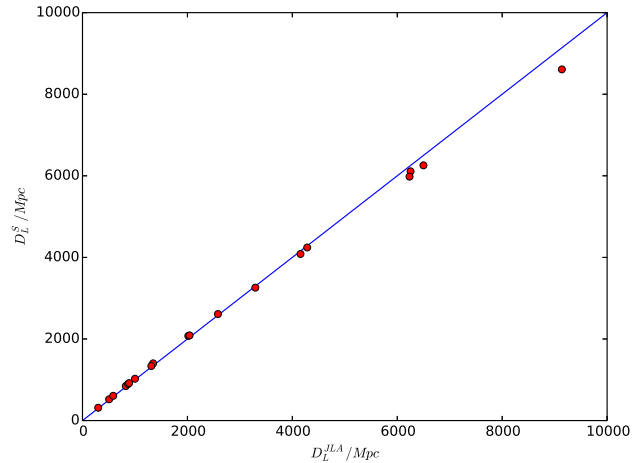}
\caption{Luminosity distances ($D_L^{JLA}$) directly from Betoule et. al. 2014 versus
         $D_L^S$. For the former, we take the best-fit parameters ($\alpha, \beta, M_B$)
         in Betoule et. al. 2014 in Eq. \ref{mub}; for the latter, we take ($\epsilon, \alpha, \beta, M_B$)
         corresponding to the minimum $\chi^2$ in Eq. \ref{dl}.}
\label{c1}
\end{figure}
\begin{figure}
\includegraphics[width=9cm]{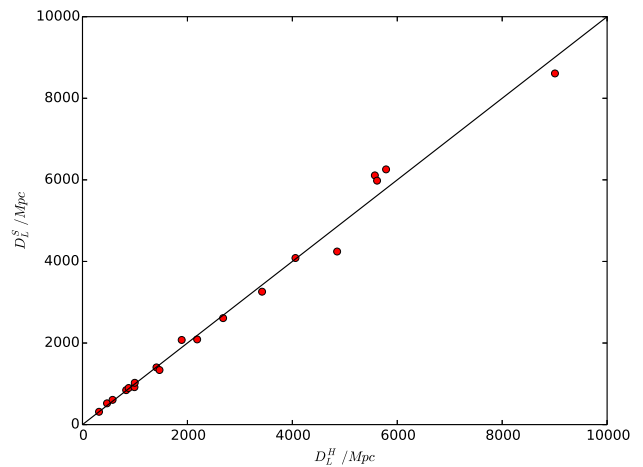}
\caption{Luminosity distances constructed from H(z) data and derived from Supernova, i.e., $D_L^H$ and $D_L^S$.
         They are consistent with each other well, corresponding to the minimum $\chi^2$.}
\label{c2}
\end{figure}

Further discussion is in order regarding our difference in standards for the SN analysis. Since Ia supernovae do not by themselves give luminosity distances, one can use either $\Lambda CDM$ or $H(z)$ to calibrate the intrinsic supernova brightness. The $\Lambda CDM$ method has been applied widely. Using $\chi^2$ statistics, it compares $\mu(\alpha, \beta, M_B)$ with $\mu^{\Lambda CDM}(\Omega_M)$ to find the best-fit for ($\alpha, \beta, M_B, \Omega_M$), thus yielding the luminosity distance $D_L(\alpha, \beta, M_B)$. In this paper, we adopt constructed luminosity distances $D_L^H$ from H(z) data as the standard, instead of the model $D_L^{\Lambda CDM}(\Omega_M)$, and so we do not have the parameter $\Omega_M$ in the standard itself. To demonstrate our method, we present Fig.~\ref{c1}, where the horizontal axis stands for the best-fit luminosity distances $D_L^{JLA}$ directly from Betoule et. al. 2014 without considering cosmic opacity, and the vertical axis stands for the derived
best-fit  luminosity distances $D_L^S$ in this paper: we first get the best-fits of ($\epsilon, \alpha, \beta, M_B$), then use Eq.~(\ref{dl}) to get the luminosity distances. The figure shows the consistency between these, especially at low redshifts where both the selection of standards and cosmic opacity
should have little impact. We also notice $D_L^S$ is slightly smaller than $D_L^{JLA}$ at large redshift. This is due to a positive
best-fit for $\epsilon$ that makes the observed image dimmer. We also compare the derived luminosity distances $D_L^S$ with the constructed ones $D_L^H$ from the $H(z)$ data in Fig.~\ref{c2} to show the relevant correction of the fitting.
\par
One of the key points in this paper was to take into account in our analysis the correlations between constructed luminosity distances from $H(z)$ data.
To show how important this is, we repeat our analysis for the case where no correlations are considered. The
resulting constraint is $\epsilon=0.044_{-0.080}^{+0.078}(1\sigma)_{-0.167}^{+0.159}(2\sigma)$ for the ¡°BEST¡± statistics
and $\epsilon=0.040_{-0.081}^{+0.077}(1\sigma)_{-0.189}^{+0.147}(2\sigma)$ for the ¡°MEAN¡± statistics. Therefore,
the impact of these correlations on the $1\sigma$ error is $\sim0.035$ and cannot be ignored. Also, with the same data, we recalculate
the constraint on $\epsilon$ using a method similar to previous studies, i.e., not taking into account these correlations and also keeping supernova parameters ($\alpha, \beta, M_B$) fixed. The result is $\epsilon=0.018\pm0.044(1\sigma)\pm0.087(2\sigma)$, corresponding to a further reduction in the $1\sigma$ error by $\sim0.035$.
Therefore, our two improvements in the analysis presented in this paper have an obvious influence on assessing the uncertainties on cosmic opacity.
Comparing to other studies in the literature, although different datasets have been used and the details in methodologies vary, previous studies (which ignored the two effects we have considered here) generally obtained a smaller limit for $\epsilon$ compared to our $1\sigma$ error $\sim0.114$. For example, in Holanda et. al. 2014 and Liao et. al. 2013, the corresponding $1\sigma$ error was found to be in the range $\sim$(0.039,0.075).
Apart from the slightly different central value of $\epsilon$, we have demonstrated that considering these correlations and allowing supernova parameters to vary are both very important for a reliable estimation of the constraint limits.
\par
We now discuss how our model-independent constraints compare to previous model-dependent bounds on cosmic opacity and their implications. The strongest model-dependent constraints on $\epsilon$ coming from distance measure comparison ($D_L$ vs $H(z)$) in the context of the DDR can be found in~\cite{Tofz,axion2} and are at the level of $\epsilon<0.04$ at 2$\sigma$. These have been derived using the Union2.1 SN compilation for $D_L$ and a collection of cosmic chronometer and BAO determinations of $H(z)$, so weak model dependences can be found both in the  $D_L$ and $H(z)$ data used, as discussed above. Our corresponding constraint derived in this paper is at the level of  $\epsilon<0.26-0.29$ (2$\sigma$), that is a factor of 6-7 weaker, but it is not dependent on assumptions on the cosmological model\footnote{In particular, as explained above, we have not used any grid in cosmological parameters but obtained constraints on $\epsilon$ through direct comparison of distance measure determinations that have been calibrated without introducing any particular cosmological model. Strictly speaking our constraint has been derived assuming a flat geometry in obtaining luminosity distances from $H(z)$ data, so in some sense we have specified the cosmological parameter $\Omega_k$. However, the independent determination of the flatness of our universe at the percent level, implies that corrections coming from deviations from flatness are of the order of $\Omega_k/6$ and are thus at least one order of magnitude smaller than the errors considered here.}.
From that point of view our constraint is remarkably strong, being of nearly the same order of magnitude as that of ~\cite{Tofz,axion2} but free of cosmological model dependences. In fact, our 1$\sigma$ limits are largely determined by the errors on the observed luminosity distances (of which we only used 19 out of the available 740 data-points), so our model-independent constraints could be further strengthened by utilising more of the available datapoints. This could be done by binning the data in bins centred around the 19 redshifts of our $H(z)$ data, or by interpolation. However, both of these could introduce systematic biases, so in this paper we have presented the cleanest and most conservative way of doing the analysis avoiding model-dependences and such systematic biases. We have thus demonstrated that strong constraints on opacity can still be obtained even if we use a small fraction of the available SN data.
\par
Finally, we consider the implications of these constraints for fundamental mechanisms that would generically give rise to cosmic opacity. A typical microphysical source of opacity would be a two-photon interaction with an unobserved particle species, which would allow astrophysical photons to decay into that species in the presence of intergalactic magnetic fields. Such interactions are typically suppressed by a high-energy physics scale, say $M$, so that the photon decay probability per unit length is small, but integrated over cosmological distances the effect can be significant, leading to observable SN dimming that could be confused with dimming due to cosmic acceleration. The best motivated example of this type is photon-axion mixing. In this case, constraints on the photon decay probability per Mpc have been obtained in~\cite{axion2} and are at the level of $P_{\rm Mpc}\simeq 4\times 10^{-5}$. From our discussion, we may expect a cosmological-model-independent constraint on $P_{\rm Mpc}$ that would be a factor of few weaker and indeed a simple computation gives:

\begin{equation}
P_{\rm Mpc}\simeq 2.5 \times 10^{-4}
\end{equation}

Subject to assumptions about the astrophysics, this can be translated into a bound on the fundamental axion-photon coupling. Since this coupling scales with the square root of $P_{\rm Mpc}$, the resulting constraint on the coupling scale $M$ is in fact only a factor of 2-3 weaker than in \cite{axion2} and is of the order of $10^{10}$ GeV, assuming magnetic fields of 1 nG coherent over domains of 1 Mpc. For chameleons, these constraints are slightly weaker as one must marginalise over an additional parameter of order unity describing how chameleons interact with matter. The effect of our analysis in this case is again to weaken the constraint on $P_{\rm Mpc}$ by a factor of few but extend its validity to any cosmological model as it is free of cosmological biases. Finally, for alternative mixing mechanisms (e.g. gravitons and Kaluza-Klein modes) mentioned in the introduction it is harder to make a quantitative connection between observational bounds on $\epsilon$ and constraints on fundamental parameters, as these effects are generally weaker and depend strongly on more model parameters, e.g. on compactification scales in the case of Kaluza-Klein modes~\cite{Deffayet2000}. From our discussion it is clear that our analysis would again imply a photon decay probability per Mpc of the order $\sim 10^{-4}$. The implications of this constraint for more fundamental parameters could be examined in a model by model basis.

\begin{table}
\begin{center}
\begin{tabular}{lcc}
\hline\hline
$$  & $BEST$ & $MEAN$  \\
\tableline
$\epsilon$ & $0.070_{-0.121}^{+0.107}(1\sigma)_{-0.253}^{+0.218}(2\sigma)$ &$0.056_{-0.122}^{+0.108}(1\sigma)_{-0.327}^{+0.198}(2\sigma)$ \\
$\alpha$ &$0.125_{-0.056}^{+0.063}(1\sigma)_{-0.114}^{+0.135}(2\sigma)$ & $0.132_{-0.055}^{+0.064}(1\sigma)_{-0.103}^{+0.177}(2\sigma)$ \\
$\beta$ &$3.266_{-0.992}^{+1.361}(1\sigma)_{-1.724}^{+3.669}(2\sigma)$ & $3.900_{-1.015}^{+2.222}(1\sigma)_{-1.486}^{+6.100}(2\sigma)$ \\
$M_B$ &$-19.132_{-0.220}^{+0.265}(1\sigma)_{-0.438}^{+0.53}(2\sigma)$ & $-19.094_{-0.227}^{+0.260}(1\sigma)_{-0.410}^{+0.640}(2\sigma)$ \\
\hline\hline
\end{tabular}
\end{center}
\caption{Best-fit values and mean values with $1\sigma$ and $2\sigma$ errors for cosmic opacity $\epsilon$
  and SN Ia nuisance parameters ($\alpha, \beta, M_B$), respectively.
}\label{stats}
\end{table}

\section{Summary}
In this paper we have presented a clean method for constraining cosmic opacity using distance measures in a model-independent way.
The motivation for independently constraining opacity becomes apparent when considering the multitude of possible sources of photon absorption
or decay of SN photons along the line of sight, which can contribute to SN dimming thus affecting the reliable reconstruction of the expansion history,
especially in the accelerated era.
Photons emitted from distant sources
might suffer from extinction by the intergalactic medium or comic dust and intervening galaxies.
Furthermore, in a wide class of theories involving two-photon interactions with other fields, photons can
decay to light axions, chameleons, gravitons or Kaluza-Klein modes in the presence of extragalactic magnetic fields.
Such mechanisms lead to an effective violation of photon number conservation, thus making the observed source
dimmer than what expected and introducing a bias in our reconstruction of universal acceleration. It is
therefore necessary and timely to quantitatively study these effects and to produce independent constraints on
cosmic opacity.
\par
There have been significant efforts recently on this topic and opacity has been constrained at the per cent level down to redshifts of $\sim 2$.
However, no study to date has achieved a completely cosmology-independent test, as the Hubble diagrams for SNe Ia used were constructed from global fits in the context of the concordance model. Moreover, most studies have ignored the correlations between different redshifts when opacity-free distances were derived from observational H(z) data. It is therefore necessary to give an improved analysis and present a clean test of cosmic opacity.
\par
To this end, we compared two kinds of luminosity distances: one
from SNe Ia, which is susceptible to cosmic opacity, and one constructed from
$H(z)$ data, which is cosmic opacity free. The SN Ia data we used were derived directly from
the measured light curves and do not depend on cosmological modelling. In addition we corrected
the inappropriate statistics used when constructing luminosity distances from $H(z)$ data in the
literature, by taking into account the correlations between different redshifts. Based on our improved
analysis, the derived constraints on opacity are somewhat weaker but the test is more robust and
more widely applicable as it does not depend on cosmological model. Our results are, as expected
from past work, consistent with a transparent universe, but our bounds can be used to constrain physical
mechanisms giving rise to opacity. As cosmological data precision improves, these methods will be
important in better quantifying opacity and accurately reconstructing dark energy parameters.

\acknowledgments
This work was supported by the Ministry of Science and Technology National Basic Science Program (Project 973) under Grants Nos.
2012CB821804 and 2014CB845806, the Strategic Priority Research Program ``The Emergence of Cosmological Structure"
of the Chinese Academy of Sciences (No. XDB09000000), the National Natural Science Foundation of China under Grants
Nos. 11373014, 11073005, 11375092 and 11435006, the China Postdoc Grant No. 2014T70043, and the Fundamental Research
Funds for the Central Universities and Scientific Research Foundation of Beijing Normal University. The work of AA was supported by
an Advanced Research Fellowship at the University of Nottingham, UK.  AA and ZL would like to thank the University of Nottingham,
China Campus, for supporting the Ningbo Cosmology Workshop ``Cosmology as a Fundamental Physics Lab" where part of this
collaboration was fostered.

\end{document}